\documentclass[conference]{IEEEtran}
\IEEEoverridecommandlockouts
\newtheorem{definition}{Definition}
\usepackage{import}
\usepackage{hyperref}

\usepackage[table]{xcolor} 
\usepackage{listings}
\usepackage{multirow}
\usepackage{amssymb}
\usepackage{cite}
\usepackage{amsmath,array,booktabs,caption,fancyvrb,graphicx,tabularx}
\usepackage{enumitem}
\usepackage{amsfonts}
\usepackage{arydshln}
\usepackage{mathtools}
\usepackage{algorithm,algpseudocode}
\usepackage{xifthen}
\usepackage{xspace}
\usepackage{textcomp}
\usepackage{pgfplots}
\usepackage{pgfplotstable}
\pgfplotsset{width=10cm,compat=1.14}
\pgfplotsset{xticklabel={\tick},scaled x ticks=false}
\pgfplotsset{plot coordinates/math parser=false}
\usepackage{subfig}
\usepackage{import}
\usepackage[normalem]{ulem}
\usetikzlibrary{patterns}
\usetikzlibrary{positioning}

\usetikzlibrary{arrows.meta}
\usetikzlibrary{graphs}
\usepackage{float}

\newcommand{\figref}[1]{Figure~\ref{fig:#1}}

\renewcommand{\eqref}[1]{Eqn.~(\ref{Eq:#1})}

\newcommand{\tableref}[1]{Table~\ref{tbl:#1}}

\newcommand{\sectref}[1]{Section~\ref{Se:#1}}

\renewcommand{\algref}[1]{Algorithm~\ref{Alg:#1}}
\newcommand{\algline}[1]{(Line~\ref{Line:#1})}
\newcommand{\alglines}[2]{(Lines~\ref{Line:#1}--\ref{Line:#2})}

\newcommand{\lstref}[1]{Listing~\ref{Lst:#1}}
\newcommand{\lstline}[1]{(Line~\ref{lstLine:#1})}

\colorlet{sjcolor}{blue}
\colorlet{ubcolor}{teal}
\colorlet{urkcolor}{red}

\newcommand{\toolname}{{\sc LLOV}\xspace}
\newcommand{\llovmhp}{{\sc LLOV (with MHP \& Polly)}\xspace}
\newcommand{\llovpolly}{{\sc LLOV (with Polly)}\xspace}
\newcommand{\llovmhponly}{{\sc LLOV (with MHP)}\xspace}
\newcommand{\ie}{{\em i.e.,}\xspace}
\newcommand{\tool}[1]{{\sc {#1}}\xspace}
\newcommand{\Omit}[1]{}
\newcommand{\EM}[1]{{\em #1}}

\newcommand{\TT}[1]{{\tt #1}}

\newcommand{\inangleb}[1]{\ensuremath{\langle#1\rangle}}
\newcommand{\insquareb}[1]{\ensuremath{[#1]}}

\newcommand{\tg}{\tool{TaskGraph}}

\newcommand{\sqrb}[1]{{$[ #1 ]$}}

\newcommand{\fixme}[2][]{\textcolor{red}{\textbf{ToDo:}\ifthenelse{\isempty{#1}{}}{}{#1--}#2}}
\newcommand\redsout{\bgroup\markoverwith{\textcolor{orange}{\rule[0.5ex]{2pt}{1pt}}}\ULon}

\DeclareMathAlphabet{\mathpzc}{OT1}{pzc}{m}{it}

\definecolor{mGreen}{rgb}{0,0.6,0}
\definecolor{mGray}{rgb}{0.5,0.5,0.5}
\definecolor{mPurple}{rgb}{0.58,0,0.82}
\definecolor{backgroundColour}{rgb}{0.95,0.95,0.92}

\lstdefinestyle{CStyle}{
    commentstyle=\color{mGreen},
    keywordstyle=\color{magenta},
    numberstyle=\tiny\color{mGray},
    stringstyle=\color{mPurple},
    basicstyle=\scriptsize,
    breakatwhitespace=false,         
    breaklines=true,                 
    captionpos=b,                    
    keepspaces=true,                 
    numbers=left,                    
    numbersep=5pt,   
    showspaces=false,                
    showstringspaces=false,
    showtabs=false,                  
    tabsize=2,
    language=C++
}

\lstdefinestyle{FortStyle}{
    commentstyle=\color{mGreen},
    keywordstyle=\color{magenta},
    numberstyle=\tiny\color{mGray},
    stringstyle=\color{mPurple},
    basicstyle=\scriptsize,
    breakatwhitespace=false,         
    breaklines=true,                 
    captionpos=b,                    
    keepspaces=true,                 
    numbers=left,                    
    numbersep=5pt,   
    showspaces=false,                
    showstringspaces=false,
    showtabs=false,                  
    tabsize=2,
    language=[90]Fortran
}

\lstdefinestyle{ompStyle}{
         language=c++, 
         basicstyle=\small,
         numbers=none,            
         numberstyle=\footnotesize,
         numbersep=5pt,            
         backgroundcolor=\color{white},
         showspaces=false,             
         showstringspaces=false,       
         showtabs=false,               
         frame=single,                 
         tabsize=4,                    
         breaklines=true,              
         columns=fullflexible,
         breakautoindent=false,
         framerule=0pt,
         xleftmargin=0pt,
         xrightmargin=0pt,
         breakindent=0pt,
         resetmargins=true,
         morekeywords={pragma,omp,parallel,sections,section,single,master,critical,atomic,simd,device},
         escapeinside={(*}{*)},
    }

\makeatletter
\lstdefinelanguage{llvm}{
  morecomment = [l]{;},
  morestring=[b]", 
  sensitive = true,
  classoffset=0,
  morekeywords={
    define, declare, global, constant,
    internal, external, private,
    linkonce, linkonce_odr, weak, weak_odr, appending,
    common, extern_weak,
    thread_local, dllimport, dllexport,
    hidden, protected, default,
    except, deplibs,
    volatile, fastcc, coldcc, cc, ccc,
    x86_stdcallcc, x86_fastcallcc,
    ptx_kernel, ptx_device,
    signext, zeroext, inreg, sret, nounwind, noreturn,
    nocapture, byval, nest, readnone, readonly, noalias, uwtable,
    inlinehint, noinline, alwaysinline, optsize, ssp, sspreq,
    noredzone, noimplicitfloat, naked, alignstack,
    module, asm, align, tail, to, 
    addrspace, section, alias, sideeffect, c, gc, 
    target, datalayout, triple,
    blockaddress
    blockaddress
  },
  classoffset=1, keywordstyle=\color{purple},
  morekeywords={
    fadd, sub, fsub, mul, fmul,
    sdiv, udiv, fdiv, srem, urem, frem,
    and, or, xor,
    icmp, fcmp,
    eq, ne, ugt, uge, ult, ule, sgt, sge, slt, sle,
    oeq, ogt, oge, olt, ole, one, ord, ueq, ugt, uge,
    ult, ule, une, uno,
    nuw, nsw, exact, inbounds,
    phi, call, select, shl, lshr, ashr, va_arg,
    trunc, zext, sext,
    fptrunc, fpext, fptoui, fptosi, uitofp, sitofp,
    ptrtoint, inttoptr, bitcast,
    ret, br, indirectbr, switch, invoke, unwind, unreachable,
    malloc, alloca, free, load, store, getelementptr,
    extractelement, insertelement, shufflevector,
    extractvalue, insertvalue,
  },
  alsoletter={\%},
  keywordsprefix={\%},
}
\makeatother

\def\BibTeX{{\rm B\kern-.05em{\sc i\kern-.025em b}\kern-.08em
    T\kern-.1667em\lower.7ex\hbox{E}\kern-.125emX}}
\begin{document}

\title{OpenMP aware MHP Analysis for Improved Static Data-Race Detection}

\author{\IEEEauthorblockN{Utpal Bora\IEEEauthorrefmark{1},
Shraiysh Vaishay\IEEEauthorrefmark{2}, 
Saurabh Joshi\IEEEauthorrefmark{3} and
Ramakrishna Upadrasta\IEEEauthorrefmark{4}
}
\IEEEauthorblockA{Computer Science and Engineering,
Indian Institute of Technology Hyderabad, India\\
Email: \{\IEEEauthorrefmark{1}cs14mtech11017,
\IEEEauthorrefmark{2}cs17btech11050\}@iith.ac.in,
\{\IEEEauthorrefmark{3}sbjoshi,
\IEEEauthorrefmark{4}ramakrishna\}@cse.iith.ac.in}
}

\maketitle
\pagestyle{plain}

\begin{abstract}
Data races, a major source of bugs in concurrent programs, can result in loss of manpower and time as well as data loss due to system failures.
OpenMP, the de facto shared memory parallelism framework used in the HPC community, also suffers from data races.
To detect race conditions in OpenMP programs and improve turnaround time and/or developer productivity, we present a data flow analysis based, fast, static data race checker in the LLVM compiler framework. 
Our tool can detect races in the presence or absence of explicit \TT{barriers}, with implicit or explicit synchronization.
In addition, our tool effectively works for the OpenMP target offloading constructs and also supports the frequently used OpenMP constructs.

We formalize and provide a data flow analysis framework to perform Phase Interval Analysis (PIA) of OpenMP programs.
Phase intervals are then used to compute the MHP (and its complement NHP) sets for the programs, which, in turn, are used to detect data races statically.

We evaluate our work using multiple OpenMP race detection benchmarks and real world applications.
Our experiments show that the checker is comparable to the state-of-the-art in various performance metrics with around 90\% accuracy, almost perfect recall, and significantly lower runtime and memory footprint.

\end{abstract}
\section{Introduction}\label{Se:intro}
Data races are a common source of bugs in parallel programs. 
Data races might result in errors as well as system failures, and these can be both fatal and/or costly. 
They are also a source of non-determinism and can be extremely difficult to reproduce.

OpenMP is the most frequently used parallel programming model/framework for High Performance Computing (HPC) within a node, and is based on shared memory parallelism.
Recently, the focus of OpenMP has been further broadened to provide support features for device offloading. This is so that it can also be used for accelerators that are increasingly adopted by the HPC community. 
Around 30\% of the top 500 supercomputers from the June 2021 list~\cite{top500url} use accelerators.

Programs written in OpenMP may suffer from data races.
Considerable amount of effort has been spent in detecting data races in OpenMP programs. 
For this purpose, both dynamic as well as (purely) static techniques have been proposed in the literature.
Most of the current OpenMP race detectors are based on dynamic analysis techniques. 

Since OpenMP internally uses POSIX threads (\TT{pthread}), the dynamic race detection tools that target \TT{pthread} based programs---such as \tool{Helgrind}\cite{valgrind2007helgrind}, \tool{Valgrind DRD}\cite{valgrind2007drd}, and \tool{TSan}\cite{Serebryany:wbia:2009}---can also naturally detect races in OpenMP programs.
However, these tools are ill-equipped to handle many high level OpenMP features, either in the form of constructs, or in the form of semantics.  Not considering these features makes these tools prone to producing many false positives.
Recent works like \tool{Archer}~\cite{Atzeni:ipdps:2016} improve the existing \TT{pthread} based tool \tool{TSan} by incorporating high-level semantics of OpenMP and subsequently improving the precision and execution time for race detection.

A major drawback of the dynamic race detection tools is that their turnaround time could be exorbitantly large. These tools are based on running the program on selected inputs. This running time is increased further because of the additional instrumentation overhead. The turnaround time for dynamic tools is also determined by its race reporting strategy.
If a tool reports only the first race and quits execution immediately, the time taken by it will be lesser when compared to another tool that reports all the possible races encountered in a particular execution.
Dynamic tools that follow the latter race reporting strategy have greater turnaround time, as some programs might take hours, or even days to complete execution.
This strategy also becomes tricky for race conditions that are dependent on the execution sequence of the threads and reproducing such bugs using dynamic techniques becomes a developer's nightmare.

Static tools, on the other hand, have the advantage of low turnaround time and also provide greater details with respect to the source location and variable name of the offending memory accesses.
However, the static race detection tools for OpenMP framework are not mature enough to provide wide coverage, and do not scale with bigger codebases.
Also most static tools such as \toolname~\cite{Bora:taco:2020}, \tool{ompVerify}~\cite{Basupalli:iwomp:2011}, \tool{PolyOMP}~\cite{Chatarasi:lcpc:2016}, and \tool{DRACO}~\cite{Ye:correctness:2018} employ only a single technique to detect data races; this method may not be adequate in supporting the multitude of features provided by OpenMP.
Hence the existing static race detection tools supports limited features of OpenMP as highlighted in \tableref{pragmas}.

In the recent work \toolname~\cite{Bora:taco:2020}, the authors used the exact polyhedral dependence analysis for race detection. Though this increases the precision of the tool, the limited scope of polyhedral tools limits the applicability of the tool itself. 
In the present work, we extend \toolname by incorporating multiple techniques (that we have described in \sectref{Implement}) to increase its coverage and improve the accuracy and recall.

In this paper, we propose a method that uses both data flow analysis techniques as well as polyhedral analysis techniques to detect races in OpenMP programs.
Our implementation is based on LLVM~\cite{Lattner:lcpc:2004} and takes race detection in parallel programs way beyond the state-of-the-art techniques~\cite{Bora:taco:2020, Swain:sc:2020}.

\textbf{Main Contributions:} In this work, we make the following contributions:\\

\begin{itemize}
    \item \EM{OpenMP aware MHP analysis}: We provide a novel May-Happen-in-Parallel (MHP) analysis technique for OpenMP programs using a new data flow analysis in LLVM.
    We compute the Never-Happen-in-Parallel (NHP) sets, the complement sets of MHP, and use this information to statically reason about possible data races present in the OpenMP programs.
    Our analysis is independent of the number of threads and their execution order, the (parametric) size of the input program, the execution paths, and the system configuration.
    
    \item \EM{Phase Interval Analysis (PIA):}
    We provide a new static analysis technique, (an adaptation of) Phase Interval Analysis (PIA), to compute the execution phases of an OpenMP program.
    The granularity of the PIA analysis is at basic block level, and it computes the minimum and the maximum phases in which a basic block can execute.
    We formalize the Phase Interval Analysis framework and provide an implementation of PIA for OpenMP programs using interval analysis over the positive integer domain.
    Our MHP analysis internally uses the execution phase intervals of the memory accesses computed by PIA.

    \item \EM{Increased Coverage:}
    Our tool \llovmhp provides support for a wider set of OpenMP constructs in comparison with 
    \toolname as highlighted in ~\tableref{pragmas}.
    In addition, we overcome the limitations of \toolname and handle the False Negative (FN) cases effectively using PIA.
        
    \item \EM{Visualization of PIA:} We provide a framework to generate \tg as DOT files to visualize phase intervals of the program. 
    Each basic block in the graph is annotated with the phase in which the program can execute.
    This helps in manually verifying the data races and improves the productivity of the parallel programmer.
    
    \item We have done an extensive evaluation of our tool on a set of race detection benchmarks commonly used in the literature and on multiple open source software with millions of lines of code demonstrating scalability of our tool.
    Our experiments show that \llovmhp is fastest and requires least amount of memory among the state-of-the-art data race checker.
    
\end{itemize}

The paper is organized in the following way: We first cover required background in~\sectref{prelim}, and we follow it up with analysis and implementation details in~\sectref{Implement}. 
We discuss related work in~\sectref{relatedWork}, and then we present experimental evaluation in~\sectref{results}, and finally we summarize in~\sectref{futurework}.
\section{Background and Motivation}\label{Se:prelim}

Parallel processing is ubiquitous in the present world.
Modern processors are mostly multiprocessors and modern operating systems support both multiprocessing as well as multithreading to provide better throughput and user experience.
To extract parallelism correctly, two programs (or a single program with multithreading) need to follow the Bernstein conditions~\cite{Bernstein:TEC:1966} for parallelism.
The program segments running in parallel must preserve the three types of data dependences -- namely, flow-, anti-, and output-dependence.  
When two program segments do not follow these conditions for parallelism, data races might occur resulting in system failure and/or non deterministic system behaviour.

\begin{definition}
	[Data Race] An execution of a concurrent program is said to have a \EM{data race} when two different threads access the same memory location, these accesses are not protected by a mutual exclusion mechanism (e.g., locks), the order of the two accesses is non-deterministic 
	and one of these accesses is a write.
\end{definition}

Using static or dynamic program analysis techniques, it is possible to detect conditions in the program that might result in data races.
Compilers for most parallel programming languages provide program analysis tools to detect these race conditions.
\toolname~\cite{Bora:taco:2020} is one such data race checker for OpenMP programs implemented in the LLVM compiler framework~\cite{Lattner:lcpc:2004}.

\toolname can statically detect data races in the most frequently used OpenMP constructs.
It uses the Polyhedral model~\cite{Feautrier:Encyclopedia:2011} to determine any violation of the dependences due to explicit parallelism.
\toolname works on the intermediate representation (IR) of LLVM and hence it can detect data races in C/C++ and FORTRAN programs.
\toolname first reconstructs the OpenMP pragmas from the IR and represents them in an in-memory representation.
Then it uses LLVM/Polly~\cite{Grosser:ppl:2012} to model the regions of code marked parallel by the OpenMP pragmas in the polyhedral representation.
\toolname then works on the reduced dependence graph (RDG) of the program segment to detect violations of dependences due to user provided explicit parallelism in a loop nest.

\toolname can detect data races in \TT{parallel}, \TT{for}, \TT{simd} pragmas with different data sharing clauses such as \TT{shared}, \TT{reduction}, \TT{private}, \TT{firstprivate}, \TT{lastprivate}, and \TT{threadprivate}.
However, \toolname has many limitations and one major drawback is in detecting races across two static control parts (SCoPs).
As shown in the example in~\lstref{NoWait}, a thread executing the \TT{parallel for}~\lstline{for_nowait} will not wait for the other threads in the team because of the \TT{nowait} clause~\lstline{ompnowait}.
Threads that have finished executing the \TT{for} loop are free to continue and execute the \TT{single} construct~\lstline{ompsingle}.
Since there is a data dependence between write to \TT{a}[i]~\lstline{ai_nowait} and read of \TT{a}[9]~\lstline{a9}, the program might result in a data race.
Such data races due to memory accesses from two different affine regions are missed by \toolname and results in False Negatives (FN).

\begin{minipage}[t]{.8\columnwidth}
\begin{lstlisting}[frame=single, style=CStyle, label=Lst:NoWait, caption={DRB013: example  program for which \toolname fails to detect the data race.}, language=C++, captionpos=b,basicstyle=\footnotesize\ttfamily, escapechar=|]
#pragma omp parallel shared(b, error) {|\label{lstLine:omppar}|
#pragma omp for nowait |\label{lstLine:ompnowait}|
    for(i = 0; i < len; i++) |\label{lstLine:for_nowait}|
      a[i] = b + a[i]*5; |\label{lstLine:ai_nowait}|
#pragma omp single |\label{lstLine:ompsingle}|
    error = a[9] + 1; |\label{lstLine:a9}|
  }
\end{lstlisting}
\end{minipage}

\toolname also does not support device offloading constructs and explicit \TT{barriers}.
In this work, we use a new static analysis technique, PIA, to overcome the limitations of \toolname.
In addition to the FN cases discussed above, we handle many more OpenMP pragmas that are not handled by \toolname as listed in~\tableref{pragmas}.

\begin{table*}[t]
\caption{Comparison of OpenMP pragma handling by OpenMP aware tools.(Y for Yes, N for No)}
\centering
\resizebox{0.97\textwidth}{!}{
%%%%%
\begin{tabular}{|l|>{\centering}m{0.09\textwidth}|>{\centering}m{0.07\textwidth}|>{\centering}m{0.07\textwidth}|>{\centering}m{0.07\textwidth}|>{\centering}m{0.07\textwidth}|>{\centering}m{0.07\textwidth}|>{\centering}m{0.07\textwidth}|>{\centering}m{0.07\textwidth}|>{\centering\arraybackslash}m{0.07\textwidth}|}
\hline
\textbf{OpenMP Pragma}         & \llovmhp             & \toolname              & \tool{OMPRacer}       & \tool{ompVerify}      & \tool{PolyOMP}        & \tool{DRACO}         & \tool{SWORD}          & \tool{Archer}         & \tool{ROMP} \\ \hline
%\midrule
\#pragma omp parallel          & \cellcolor{green!25}Y & \cellcolor{green!25}Y & \cellcolor{green!25}Y & \cellcolor{green!25}Y & \cellcolor{green!25}Y & \cellcolor{green!25}Y & \cellcolor{green!25}Y & \cellcolor{green!25}Y & \cellcolor{green!25}Y\\
\#pragma omp for               & \cellcolor{green!25}Y & \cellcolor{green!25}Y & \cellcolor{green!25}Y & \cellcolor{green!25}Y & \cellcolor{green!25}Y & \cellcolor{green!25}Y & \cellcolor{green!25}Y & \cellcolor{green!25}Y & \cellcolor{green!25}Y\\
\#pragma omp parallel for      & \cellcolor{green!25}Y & \cellcolor{green!25}Y & \cellcolor{green!25}Y & \cellcolor{green!25}Y & \cellcolor{green!25}Y & \cellcolor{green!25}Y & \cellcolor{green!25}Y & \cellcolor{green!25}Y & \cellcolor{green!25}Y\\
\#pragma omp critical          & \cellcolor{green!25}Y   & \cellcolor{red!25}N & \cellcolor{green!25}Y & \cellcolor{red!25}N   & \cellcolor{red!25}N   & \cellcolor{red!25}N   & \cellcolor{green!25}Y & \cellcolor{green!25}Y & \cellcolor{green!25}Y\\
\#pragma omp atomic            & \cellcolor{green!25}Y   & \cellcolor{red!25}N & \cellcolor{green!25}Y & \cellcolor{red!25}N   & \cellcolor{red!25}N   & \cellcolor{red!25}N   & \cellcolor{green!25}Y & \cellcolor{green!25}Y & \cellcolor{green!25}Y\\
\#pragma omp master            & \cellcolor{green!25}Y   & \cellcolor{red!25}N & \cellcolor{green!25}Y & \cellcolor{red!25}N   & \cellcolor{green!25}Y & \cellcolor{red!25}N   & \cellcolor{green!25}Y & \cellcolor{green!25}Y & \cellcolor{green!25}Y\\
\#pragma omp single            & \cellcolor{green!25}Y   & \cellcolor{red!25}N & \cellcolor{green!25}Y & \cellcolor{red!25}N   & \cellcolor{green!25}Y & \cellcolor{red!25}N   & \cellcolor{green!25}Y & \cellcolor{green!25}Y & \cellcolor{green!25}Y\\
\#pragma omp simd              & \cellcolor{green!25}Y & \cellcolor{green!25}Y & \cellcolor{green!25}Y & \cellcolor{red!25}N   & \cellcolor{red!25}N   & \cellcolor{green!25}Y & \cellcolor{red!25}N   & \cellcolor{red!25}N   & \cellcolor{red!25}N\\
\#pragma omp parallel for simd & \cellcolor{green!25}Y & \cellcolor{green!25}Y & \cellcolor{green!25}Y & \cellcolor{red!25}N   & \cellcolor{red!25}N   & \cellcolor{green!25}Y & \cellcolor{red!25}N   & \cellcolor{red!25}N   & \cellcolor{red!25}N\\
\#pragma omp parallel sections & \cellcolor{green!25}Y   & \cellcolor{red!25}N & \cellcolor{red!25}N & \cellcolor{red!25}N   & \cellcolor{red!25}N   & \cellcolor{red!25}N   & \cellcolor{green!25}Y & \cellcolor{green!25}Y & \cellcolor{green!25}Y\\
\#pragma omp sections          & \cellcolor{green!25}Y   & \cellcolor{red!25}N & \cellcolor{red!25}N & \cellcolor{red!25}N   & \cellcolor{red!25}N   & \cellcolor{red!25}N   & \cellcolor{green!25}Y & \cellcolor{green!25}Y & \cellcolor{green!25}Y\\
\#pragma omp threadprivate     & \cellcolor{green!25}Y & \cellcolor{green!25}Y & \cellcolor{green!25}Y & \cellcolor{red!25}N   & \cellcolor{red!25}N   & \cellcolor{red!25}N   & \cellcolor{red!25}N   & \cellcolor{green!25}Y & \cellcolor{green!25}Y\\
\#pragma omp ordered           & \cellcolor{green!25}Y & \cellcolor{green!25}Y & \cellcolor{red!25}N & \cellcolor{red!25}N   & \cellcolor{red!25}N   & \cellcolor{red!25}N   & \cellcolor{red!25}N   & \cellcolor{green!25}Y & \cellcolor{green!25}Y\\
\#pragma omp distribute        & \cellcolor{green!25}Y & \cellcolor{green!25}Y & \cellcolor{green!25}Y & \cellcolor{red!25}N   & \cellcolor{red!25}N   & \cellcolor{red!25}N   & \cellcolor{red!25}N   & \cellcolor{green!25}Y & \cellcolor{green!25}Y\\
\#pragma omp task              & \cellcolor{red!25}N   & \cellcolor{red!25}N & \cellcolor{red!25}N & \cellcolor{red!25}N   & \cellcolor{red!25}N   & \cellcolor{red!25}N   & \cellcolor{red!25}N   & \cellcolor{green!25}Y & \cellcolor{green!25}Y\\
\#pragma omp taskgroup         & \cellcolor{red!25}N   & \cellcolor{red!25}N & \cellcolor{red!25}N & \cellcolor{red!25}N   & \cellcolor{red!25}N   & \cellcolor{red!25}N   & \cellcolor{red!25}N   & \cellcolor{green!25}Y & \cellcolor{green!25}Y\\
\#pragma omp taskloop          & \cellcolor{red!25}N   & \cellcolor{red!25}N & \cellcolor{red!25}N & \cellcolor{red!25}N   & \cellcolor{red!25}N   & \cellcolor{red!25}N   & \cellcolor{red!25}N   & \cellcolor{green!25}Y & \cellcolor{green!25}Y\\
\#pragma omp taskwait          & \cellcolor{red!25}N   & \cellcolor{red!25}N & \cellcolor{red!25}N & \cellcolor{red!25}N   & \cellcolor{red!25}N   & \cellcolor{red!25}N   & \cellcolor{red!25}N   & \cellcolor{green!25}Y & \cellcolor{green!25}Y\\
\#pragma omp barrier           & \cellcolor{green!25}Y   & \cellcolor{red!25}N & \cellcolor{green!25}Y & \cellcolor{red!25}N   & \cellcolor{green!25}Y & \cellcolor{red!25}N   & \cellcolor{green!25}Y & \cellcolor{green!25}Y & \cellcolor{green!25}Y\\
\#pragma omp teams             & \cellcolor{green!25}Y   & \cellcolor{red!25}N & \cellcolor{green!25}Y & \cellcolor{red!25}N   & \cellcolor{red!25}N   & \cellcolor{red!25}N   & \cellcolor{red!25}N   & \cellcolor{red!25}N   & \cellcolor{red!25}N\\
\#pragma omp target            & \cellcolor{green!25}Y   & \cellcolor{red!25}N & \cellcolor{green!25}Y & \cellcolor{red!25}N   & \cellcolor{red!25}N   & \cellcolor{red!25}N   & \cellcolor{red!25}N   & \cellcolor{red!25}N   & \cellcolor{red!25}N\\
\hline
%%%%%
\end{tabular}
}
\label{tbl:pragmas}
\end{table*}
\section{Analysis and Implementation}\label{Se:Implement}

We have used LLVM-12.0~\cite{Lattner:cgo:2004} compiler infrastructure to implement the new static analyses.
\toolname workflow is divided into two phases, analysis and verification.
The analysis phase collects OpenMP pragmas from LLVM IR and represents them in an \textit{in-memory representation} described in our prior work~\cite{Bora:taco:2020}.
In the current work, we add support for the \TT{barriers} and device offloading pragmas with a focus on the verification phase.

In the verification phase, we construct a static TaskGraph from the LLVM IR and the directives constructed in the analysis phase of \toolname.
In the following section, we provide a formal definition, an algorithm, and details about the TaskGraph.

\subsection{Task Graph}\label{SubSe:taskGraph}

\tikzstyle{multiver}=[circle,fill=black!30,minimum size=20pt,inner sep=0pt]
\tikzstyle{vertex}=[circle,fill=black!10,minimum size=20pt,inner sep=0pt]

\begin{definition}[Reduced \tg]
	A reduced \tg $G=\inangleb{V,E,R,T}$ is defined as follows: it consists of a vertex set $V$ which denotes implicit and explicit tasks. Also, its edge set $E \subseteq V\times V$ contains directed edges, a special vertex $R$ for the root of the graph, and a special sentinel vertex $T$ that represents the termination of all the tasks.
\end{definition}

A vertex $v\in V$ 
	is either a statement or a basic block in the program. The \TT{barriers} are modeled as standalone vertices, \ie a basic block will not contain more than one \TT{barrier}.

	A directed edge $e\in E$ where $e=(u,v)$ denotes that  
	$v$ can be reached directly from $u$ in some execution of the program.

Since a node is a static representation of a program statement/basic block, it can give rise to multiple instances during an execution.
\begin{definition}[Phase]
The execution of instances of nodes in the \tg happens in phases. 
	Instances of two nodes $(u,v) \in G$ will execute in two different phases of execution when there exists an implicit or explicit \TT{barrier} separating $u$ and $v$ and one of them dominates the other in the CFG. i.e., each thread executing the nodes must cross the \TT{barrier} in between the nodes. A phase is represented by a positive integer $p \in \mathbb{N}^+\cup \infty$.
\end{definition}

\begin{definition}[Phase Interval]
	To conservatively capture in which phase any instance of $u \in G$ would execute, we associate with it a phase interval \insquareb{lb,ub}, where
	$lb$ represents the least phase and $ub$ represents the maximum phase in which some instance of the node $u$ can be executed at runtime.
\end{definition}

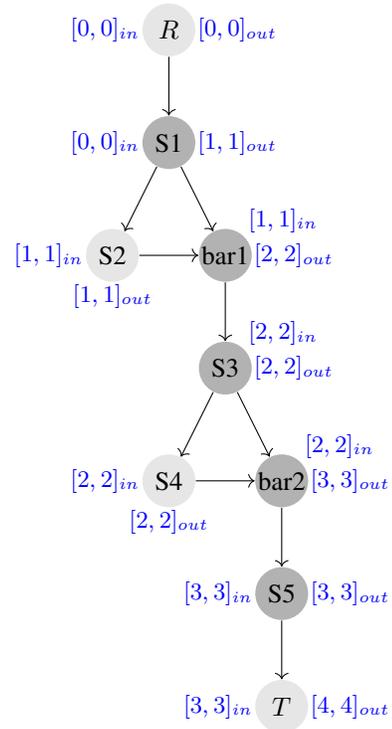
\begin{figure}
\centering
\begin{tikzpicture}[ ->,label distance=-1mm, every label/.append style={font=\small,text=blue}]
\centering
  \node[vertex,label=right:{\sqrb{0,0}$_{out}$}, label=left:{\sqrb{0,0}$_{in}$}]{$R$}
    child { node[multiver,label=right:{\sqrb{1,1}$_{out}$}, label=left:{\sqrb{0,0}$_{in}$}] (S1) {S1} 
      child { node[vertex,label=south:{\sqrb{1,1}$_{out}$}, label=left:{\sqrb{1,1}$_{in}$}] (S2) {S2} 
      edge from parent node[left,blue] {} }
      child { node[multiver,label=right:{\sqrb{2,2}$_{out}$}, label=north east:{\sqrb{1,1}$_{in}$}] (bar1) {bar1}
        child { node[multiver,label=right:{\sqrb{2,2}$_{out}$}, label=north east:{\sqrb{2,2}$_{in}$}] (S3) {S3} 
          child { node[vertex,label=south:{\sqrb{2,2}$_{out}$}, label=left:{\sqrb{2,2}$_{in}$}] (S4) {S4} 
          edge from parent node[left,blue] {} }
          child { node[multiver,label=right:{\sqrb{3,3}$_{out}$}, label=north east:{\sqrb{2,2}$_{in}$}] (bar2) {bar2} 
            child { node[multiver,label=right:{\sqrb{3,3}$_{out}$}, label=left:{\sqrb{3,3}$_{in}$}] (S5) {S5}
            child { node[vertex,label=right:{\sqrb{4,4}$_{out}$}, label=left:{\sqrb{3,3}$_{in}$}] {$T$}
          edge from parent node[left,blue] {} }
          edge from parent node[left,blue] {} }
          edge from parent node[left,blue] {} }}}};
  \path (S2) edge node[below,blue] {} (bar1);
  \path (S4) edge node[below,blue] {} (bar2);
\end{tikzpicture}
\caption{\tg for the OpenMP \TT{single} construct annotated with phase intervals.}
\label{fig:ompParallelSingle2}
\end{figure}

Unlike other task graphs used in the literature where each task is modelled separately~\cite{Gu:sc:2018} or, two instance per task is considered statically~\cite{Barik:lcpc:2005}, we  model each task only once and the \tg nodes have one-to-one correspondence with the CFG nodes except for the two special nodes $R$ and $T$.
We then use OpenMP semantics to determine nodes of the \tg that can be executed by more than one thread.
A node $u \in G$ can either have only one instance during execution or have multiple instances.
Nodes having a single execution instance are marked in light grey while nodes with multiple execution instances are marked in dark grey as shown in~\figref{ompParallelSingle2}.

\textbf{\subsection{Task Graph Construction}} \label{SubSe:taskGraphConst}
The \tg is constructed from the CFG of the program, call graph, and OpenMP semantics.
Each node in the \tg corresponds to a basic block in the CFG.
In addition to the control flow edges, the \tg also has the function call edges.
The \tg also contains edges for OpenMP semantics. 
For example, for the \TT{single} construct~\lstline{single1}  in~\lstref{ompParallelSingle2}, there will be an edge from the immediate dominator $S1$~\lstline{s1} of the \TT{single} construct to the immediate post-dominator implicit \TT{barrier}~\lstline{bar1} of the \TT{single} construct as all but one of the threads will bypass the \TT{single} construct. 
This can be seen in~\figref{ompParallelSingle2} that there exists an edge from node \TT{S1} to node \TT{bar1}.
Similarly, there exists an edge from node \TT{S3} to node \TT{bar2}.

Each node in the \tg is labelled with a phase interval, for example
a phase interval \sqrb{lb,ub} for node $S1$ denotes that $lb$ is the minimum phase and $ub$ is the maximum phase in which the node $S1$ can execute.

\EM{Phase Change:} Phase change happens only in one of the following circumstances.
Upon entering a parallel region (node \TT{S1} in~\figref{ompParallelSingle2}), leaving a parallel region (node \TT{T} in~\figref{ompParallelSingle2}), or encountering a \TT{barrier} (nodes \TT{bar1, bar2} in~\figref{ompParallelSingle2}).
The phase change can occur because of an explicit \TT{barrier} or implicit \TT{barrier} present in the constructs- \TT{parallel}, \TT{for}, \TT{workshare}, \TT{single}, \TT{sections}, \TT{taskwait}, \TT{taskgroup}, and \TT{taskloop}.

The phase change for the OpenMP pseudocode in~\lstref{ompParallelSingle2} is illustrated in~\figref{ompParallelSingle2}.
Each node is annotated with the incoming phase interval and the outgoing phase interval, which is computed by the transfer function (discussed later) on the node.

\begin{minipage}{.4\textwidth}
\begin{lstlisting}[frame=single, style=CStyle, label=Lst:ompParallelSingle2, caption={Multiple OpenMP \TT{single} constructs}, language=C++, captionpos=b,basicstyle=\scriptsize\ttfamily, escapechar=|]
#pragma omp parallel
  {
    // S1 |\label{lstLine:s1}|
#pragma omp single |\label{lstLine:single1}|
    {
      // S2 |\label{lstLine:s2}|
    } // implicit barrier |\label{lstLine:bar1}|
    // S3 |\label{lstLine:s3}|
#pragma omp single |\label{lstLine:single2}|
    {
      // S4 |\label{lstLine:s4}|
    } // implicit barrier |\label{lstLine:bar2}|
    // S5 |\label{lstLine:s5}|
  }
\end{lstlisting}
\end{minipage}\hfill

\subsection{Phase Interval Analysis}\label{SubSe:PIA}

We adapt Phase Interval Analysis (PIA)~\cite{Joshi:ipdpsw:2012} for OpenMP programs using a data flow analysis framework. We propose a forward flow analysis to compute phase intervals.
Each statement/basic-block is assigned a phase interval comprising of the minimum and the maximum phase of execution.
We perform a modified interval analysis, where the domain of the analysis is the interval lattice, and the particular abstract domain is the interval polyhedra~\cite{Cousot:ISP:1976}.

Here we formally define the data flow framework for Phase Interval Analysis.

\noindent Let,  $(I,\sqsubseteq,\sqcup,\sqcap)$ is a lattice over an interval domain. 
	An interval $PI \in I$ is of the form $\insquareb{lb,ub}$, where $lb,ub \in \mathbb{N}\cup\{0,\infty\}$ and  $0 \leq lb \leq ub \leq \infty$.
	For this lattice, $\top = \insquareb{0,\infty}$. Any interval $\insquareb{lb,ub}$ represents $\bot$ whenever $lb > ub$.
	For two phase intervals $PI_1=\insquareb{lb_1,ub_1}$ and $PI_2=\insquareb{lb_2,ub_2}$,  $PI_1 \sqsubseteq PI_2$ if an only if $lb_2 \leq lb_1 \leq ub_1 \leq ub_2$. Join ($\sqcup$)  of two phase intervals $PI_1=\insquareb{lb_1,lb_2}$
	and $PI_2=\insquareb{lb_2,ub_2}$ is defined as $\insquareb{\min(lb_1,lb_2),\max(ub_1,ub_2)}$. Similarly, meet ($\sqcap$) of two phase intervals $PI_1=\insquareb{lb_1,lb_2}$
	and $PI_2=\insquareb{lb_2,ub_2}$ is defined as $\insquareb{\max(lb_1,lb_2),\min(ub_1,ub_2)}$. We also define multiplication of a non-negative integer $c$ with $PI=\insquareb{lb,ub}$
	as $c*PI = \insquareb{c*lb,c*ub}$. Difference between two intervals $PI_1=\insquareb{lb_1,ub_1}$ and $PI_2=\insquareb{lb_2,ub_2}$ is defined as $PI_1\Delta PI_2 = \insquareb{lb_2-lb_1,ub_2-ub_1}$. First and second component of an interval $PI$ can be accessed via $PI.first$ and $PI.second$ respectively.

\indent \textbf{Data flow Equation} for Phase Interval Analysis for a node $S \in G$ is defined as
    \[
    PI_{in}[S] = \bigsqcup\limits_{p \in pred[S]} PI_{out}[p] 
    \]
    where the confluence operator is the join operation ($\sqcup$), $p$ is an immediate predecessor node of $S$, and $pred[S]$ is the set of all the immediate predecessor nodes of $S \in G$.

\indent \textbf{Transfer Function} is defined as 
    $PI_{out}[S]=f(PI_{in}[S])$
    where $f$ is identity function if the basic block $S$ does not have a \TT{barrier} and,
    $f$ is $PI_{out}(S) = PI_{in}(S) + [1,1]$
    if the basic block $S$ has a \TT{barrier} or $S$ either is starting or exiting block of a parallel construct.
    The binary operator $+$ is defined as 
    $[lb_1, ub_1] + [lb_2, ub_2] = [lb_1 + lb_2, ub_1 + ub_2]$.

    $    PI_{out}(S) =
        f(PI_{in}(S)) $
    \[
    f(PI_{in}(S)) =
    \begin{cases}
        PI_{in}(S), & \text{Identity function, if } S \\
        & \text{does not have a \TT{barrier}}.\\     
        PI_{in}(S) + [1,1], & \text{otherwise.} 
    \end{cases}
   \]

    A basic block will have at the most one \TT{barrier} (at the end/beginning). In the case of multiple \TT{barriers} in a basic block, a preprocessing step can split such basic blocks to ensure up to one \TT{barrier} per basic block.
    
    \textbf{Initialization}: All the nodes $u \in G$ are initialized with $PI_{in}[u] = \bot$ and $PI_{out}[u]=\bot$.
    The root node $R$ is initialized with phase interval $PI_{in}[R] = PI_{out}[R] = [0,0]$.

\indent  \textbf{Widening}:
    Since the interval lattice has infinite elements as well as infinite levels, we need to put a restriction on the height to make the framework terminating.
    Therefore, we represent $\infty$ with INT\_MAX in our implementation.
    However, this can be changed with the flag \textit{openmp-pia-lattice-upper-bound}.
    To reduce time complexity, we apply widening with thresholds using the loop bounds of loops for which the trip count is known. Otherwise widening happens with the default upper bound.\\
    We define widening ($\nabla_{\insquareb{lbt,ubt}}$) with thresholds $lbt$ and $ubt$ between two intervals $PI_1$ and $PI_2$ as $PI_1 \nabla_{\insquareb{lbt,ubt}} PI_2 = \insquareb{lbw,ubw}$, where $lbw=lbt$ if $lb_2<lb_1$, otherwise $lbw=lb_1$, and $ubw=ubt$ if $ub_2>ub_1$, otherwise $ubw=ub_1$.

  For a loop header $H$, let $PI ^1 _{in}(H)$ denote the phase interval after the loop body is iterated once by the phase interval analysis. If the trip-count $TC$ is known for a loop,
  we can safely accelerate computation of $PI_{in}(H) = PI^0_{in}(H) \sqcup (PI^0_{in}(H)+(TC*(PI^0_{in}(H) \Delta PI^1_{in}(H))))$ to capture the effect of the analysis going around the loop $TC$ times. If the trip count $TC$ is not known then $PI_{in}(H) = PI^0_{in}(H) \nabla_{\insquareb{0,\infty}} PI^1_{in}(H)$.

\indent  \textbf{Termination}: The transfer function used for interval analysis is monotone, as it always preserves or increases the interval level in the interval lattice.
The flow analysis is guaranteed to terminate~\cite{Cousot:ISP:1976} due to the widening operation described above.

\algref{PIA} demonstrates the phase interval computation using data flow analysis.
It takes the \tg $G$ as input and returns the phase intervals of all the nodes in $G$.

We follow a standard iterative data flow worklist algorithm where the root node is initialized with the interval \insquareb{0,0}~\algline{piainitroot}.
The algorithm runs until the worklist is empty~\alglines{piawhile}{piaendwhile}, that is, when the framework converges. 
\algline{piajoin} performs the join operation on the phase intervals of all the immediate predecessors of a node.
Widening \algline{piawiden} is applied on a loop header when the phase change $\Delta(PI)$ in one iteration of the loop could be determined.
The transfer operation \algline{piatransfer} always preserves or increases the phase interval in the interval lattice.

\begin{algorithm}
\caption{Phase Interval Analysis\label{Alg:PIA}}
\begin{algorithmic}[1]
\State \textbf{Input:} \tg $G$
\State \textbf{Output:} Phase Intervals $PI$
\State \textbf{Data:} worklist
\Function{PIA}{$G$}
  \State worklist.add($G.root$)
  \ForAll{n $\in $ G} \label{Line:piainit} 
     \State $PI_{out}[n] \gets \bot$ \Comment{init to bot}
     \State \algorithmicif\ {$n$ has \TT{barrier}} \algorithmicthen\ {worklist.add($n$)} \algorithmicend\ \algorithmicif
  \EndFor \label{Line:piaendinit}
  \State $PI_{in}[G.root] \gets [0,0]$ \label{Line:piainitroot}

  \While{!worklist.empty()} \label{Line:piawhile}
     \State $n \gets $ worklist.pop()
     \State $PI_{in}[n] \gets \bigsqcup\limits_{p \in pred[n]} PI_{out}[p]$ \Comment{join} \label{Line:piajoin}
     \If{$n$ is loop header}
        \State $PI_{out}[n] \gets w(PI_{in}[n])$ \Comment{widen} \label{Line:piawiden}
     \EndIf
     \State $PI_{out}[n] \gets f(PI_{in}[n])$ \Comment{transfer} \label{Line:piatransfer}
     \If{$PI_{out}[n] \neq PI_{in}[n]$}
        \ForAll{$s \in \text{successor}(n)$}
          \State worklist.add($s$)
        \EndFor
     \EndIf
  \EndWhile \label{Line:piaendwhile}
  \State \Return $PI_{out}$
 \EndFunction
\end{algorithmic}
\end{algorithm}

\subsection{May-Happen-in-Parallel Analysis}\label{SubSe:MHP}
The PIA described in the above section is used to compute the Never-Happen-in-Parallel (NHP) and its complement May-Happen-in-Parallel (MHP) sets.

We define the binary operator $\bigcap$ on the phase intervals $PI_1=\insquareb{lb_1,ub_1}$ and $PI_2=\insquareb{lb_2,ub_2}$ as an overlapping function, which returns true if the two intervals overlap partially or fully, and returns false otherwise.
\[
PI_1 \bigcap PI_2 = 
    \begin{cases}
      true, & lb_1 \leq ub_2 \leq ub_1 \lor \\   & lb_2 \leq ub_1 \leq ub_2 \\   
      false, & \text{otherwise.}
    \end{cases}
\]

Instances of two nodes $u, v \in G$ may run in parallel if and only if $PI[u] \bigcap PI[v]$.

\EM{Special Case:} The \TT{critical} construct needed special treatment since its scope is global. i.e., two \TT{critical} constructs in the global scope can not run in parallel since they share the same lock variable underneath.
Therefore, two \TT{critical} constructs will never run in parallel if they share the same lock.
Similarly, the \TT{master} construct also needs a special treatment. Two \TT{master} constructs within a \TT{parallel} region will not run in parallel even though they have same phase interval.

\subsection{Race Detection}\label{SubSe:RaceDetection}

Once the phase intervals are computed using PIA, potential data races can be detected by checking if the phase intervals of two memory accesses can overlap or not.
This means, the source memory access $\TT{MA_1}$ (in basic block $u$) and sink memory access $\TT{MA_2}$ (in basic block $v$) of a data dependence may potentially race if $PI[u] \bigcap PI[v]$.

\llovmhp reports data races with the precise source location of the offending memory accesses. In \lstref{ErrorReport}, we show a data race reported by \llovmhp. It also highlights the source code with line numbers. 

\begin{minipage}{.95\columnwidth}
\begin{lstlisting}[frame=single, style=CStyle, label=Lst:ErrorReport, caption={Error reporting:  \llovmhp highlights the source location of the race condition.}, language=C++, captionpos=b,basicstyle=\scriptsize\ttfamily, numbers=none, escapechar=|]
Data Race detected.
Source : llvm/lib/Transforms/OpenMPVerify/test/10.race1.c:10:11
Sink : llvm/lib/Transforms/OpenMPVerify/test/10.race1.c:12:11
==============
9 :     {
|\textcolor{red}{10 :       { x = 1; }}|
11 : #pragma omp section
|\textcolor{red}{12 :       { x = 2; }}|
13 :     }
==============
\end{lstlisting}
\end{minipage}\hfill

\subsection{Limitations}\label{SubSe:Limit}
Our PIA analysis in \llovmhp does not yet support tasking constructs (as listed in~\tableref{pragmas}). 
Although there is no fundamental limitation of the technique to provide support for tasking constructs, we did not explore this aspect in this work.
We plan to explore them in the future.
Also, the current version of \llovmhp does not perform lockset analysis~\cite{Savage:tocs:1997} and hence might produce False Positives for memory accesses guarded by exclusive locks.

\section{Related Work}
\label{Se:relatedWork}

The problem of MHP was first studied by Naumovich et al.~\cite{Naumovich:fse:1998,Naumovich:fse:1999} for X10 and Java programs.
Their technique involves modeling the program as parallel execution graph (PEG) with one CFG per thread, whereas our approach is for OpenMP programs and the \tg is independent of the number of runtime threads.
Also, they use MHP sets, while our work involves performing phase interval analysis.

Joshi et al.~\cite{Joshi:ipdpsw:2012} extended the work of Agarwal et al.~\cite{Agarwal:ppopp:2007} for programs with dynamic \TT{barriers} and introduced Phase Interval Analysis (PIA) for X10 programs.
We adapt PIA for OpenMP semantics and introduce the monotone framework based interval analysis to compute MHP information.
Recent works~\cite{Saha:ppopp:2020,Sankar:cc:2016} have improved the algorithm by Agarwal et al.~\cite{Agarwal:ppopp:2007} for X10 programs.

MHP for OpenMP programs was first explored by Chatarasi et al.~\cite{Chatarasi:lcpc:2016} using an extended polyhedral framework.
Our approach of using interval analysis based on data flow analysis framework is different from the polyhedral phase mapping of the statements incorporated by them.

Extensive work has gone into data race detection in parallel programs using different techniques such as locket analysis based~\cite{Savage:tocs:1997,Engler:sosp:2003,Yu:sosp:2005,Kasikci:hotdep:2012,Voung:fse:2007,Pratikakis:pldi:2006}, happens-before~\cite{Lamport:cacm:1978} relation based~\cite{valgrind2007drd,valgrind2007helgrind,Flanagan:pldi:2009,Serebryany:wbia:2009}, constraint solver based~\cite{Chatarasi:impact:2016}, offset span label based~\cite{Mellor-Crummey:sc:1991,Gu:sc:2018}, polyhedral model based~\cite{Bora:taco:2020,Ye:correctness:2018,Chatarasi:lcpc:2016,Basupalli:iwomp:2011}, and MHP information based~\cite{Barik:lcpc:2005,Joshi:ipdpsw:2012,Chatarasi:lcpc:2016}.
A majority of these techniques are either POSIX thread~\cite{posix2017} based, or are specific to race detection in explicit parallelism of various programming languages, such as Java, C\#, X10, and Chapel.

Race detection in OpenMP parallel programs has been studied recently in \tool{LLOV}~\cite{Bora:taco:2020}, \tool{OMPRacer}~\cite{Swain:sc:2020}, \tool{SWORD}~\cite{Atzeni:ipdps:2018}, \tool{ROMP}~\cite{Gu:sc:2018}, \tool{DRACO}~\cite{Ye:correctness:2018}, \tool{Archer}~\cite{Atzeni:ipdps:2016}, \tool{PolyOMP}~\cite{Chatarasi:lcpc:2016}, and \tool{ompVerify}~\cite{Basupalli:iwomp:2011}.
Race detection tools for POSIX threads based parallel programs such as \tool{TSan-LLVM}~\cite{Serebryany:rv:2011}, \tool{Helgrind}~\cite{valgrind2007helgrind}, and \tool{Valgrind DRD}~\cite{valgrind2007drd} can also analyze OpenMP programs as OpenMP uses \TT{pthread} underneath it.

We limit the discussion to the state-of-the-art tools that we compare against.

\tool{OMPRacer}~\cite{Swain:sc:2020} is a static analysis tool that can detect data races in OpenMP programs.
\tool{OMPRacer} employs alias analysis, value-flow analysis using scalar evaluations, and static happens-before graph to detect data races.
It constructs the static happens-before graph for all the memory operations and models threads using two nodes per operation in the graph.
\tool{OMPRacer} does not generate any warning for pragmas, such as \TT{task}, \TT{sections}, \TT{ordered}, \TT{doacross}, etc., that are not supported by it.

\tool{Archer}~\cite{Atzeni:ipdps:2016} uses both static and dynamic analyses for race detection. 
It uses happens-before relations which enforces multiple runs of the program to find races. 
\tool{Archer} reduces the analysis space of \TT{pthread} based tool \tool{TSan-LLVM} by instrumenting only parallel sections of an OpenMP program.
\tool{Archer} uses OMPT~\cite{Eichenberger:iwomp:2013} callbacks to instrument the code with happens-before annotations for \tool{TSan-LLVM}.

\tool{SWORD}~\cite{Atzeni:ipdps:2018} is a dynamic tool based on operational semantic rules and uses OpenMP tools framework OMPT~\cite{Eichenberger:iwomp:2013}.
\tool{SWORD} uses locksets to implement the semantic rules by taking advantage of the events tracked by OMPT.
\tool{SWORD} cannot detect races in OpenMP \TT{simd}, \TT{task} and target offloading constructs.

\tool{ROMP}~\cite{Gu:sc:2018} is a dynamic data race detection tool for OpenMP programs which maintains access history for memory accesses. 
\tool{ROMP} builds upon the offset-span-labels of OpenMP threads and constructs task graphs to detect races.

\tool{Helgrind}~\cite{valgrind2007helgrind} is a dynamic data race detection tool in the Valgrind framework~\cite{valgrind2003url} for C/C++ multithreaded programs.
\tool{Helgrind} maintains \textit{happens-before} relations for each pair of memory accesses and forms a directed acyclic graph (DAG). 

\tool{Valgrind DRD}~\cite{valgrind2007drd} is another dynamic race detection tool in Valgrind. 
It can detect races in multithreaded C/C++ programs.
It is based on happens-before relations similar to \tool{Helgrind}.

\tool{TSan-LLVM}~\cite{Serebryany:rv:2011} is a dynamic tool based on \tool{ThreadSanitizer}~\cite{Serebryany:wbia:2009}.
\tool{TSan-LLVM} uses LLVM to instrument the binaries in place of Valgrind.
\tool{TSan-LLVM} instrumented binaries incur less runtime overhead compared to \tool{ThreadSanitizer}. 
However, it still has similar memory requirements and remains a bottleneck for larger programs. 

Tools such as 
\tool{Relay}~\cite{Voung:fse:2007}, 
\tool{Locksmith}~\cite{Pratikakis:toplas:2011}, and
\tool{RacerX}~\cite{Engler:sosp:2003}
use \tool{Eraser}'s~\cite{Savage:tocs:1997} lockset analysis algorithm and statically detect races in \TT{pthread} based C/C++ programs.
Other tools such as \tool{Eraser}~\cite{Savage:tocs:1997}, \tool{FastTrack}~\cite{Flanagan:pldi:2009}, \tool{CoRD}~\cite{Kasikci:hotdep:2012}, and \tool{RaceTrack}~\cite{Yu:sosp:2005} use the lockset algorithm to dynamically detect races in parallel programs.
Since they are not specific to OpenMP, we have not discussed them here. \\
\section{Experimental Setup \& Results}\label{Se:results}

In this section, we evaluate our approach on a set of standard data race benchmarks commonly used in the literature. We also study the scalability of our approach on a few real world applications adding up to a million lines of code that use OpenMP for shared memory parallelism.

We compare our work in three different modes- 1) in \llovmhponly, we enable our new MHP verification engine, 2) in \llovpolly, we enable only Polyhedral verification, and 3) in \llovmhp, we enable both MHP and Polyhedral verification.
In mode 2 and 3, alias analysis checks are disabled whereas they were enabled by default in \toolname~\cite{Bora:taco:2020}.

Our toolchain is setup such that both MHP and Polyhedral verification engines are enabled by default. This is same as \llovmhp.
However, one can disable each verification engine using command line flags.
Polyhedral verification engine can be disabled with \EM{openmp-verify-use-polly-da=false} and MHP verification engine can be disabled with flag \EM{openmp-verify-use-llvm-da=false}.

\textbf{System Configuration: }
Our experimental setup consists of a server with two Intel Xeon E$5$-$2697$ v$4$ processors having a clock frequency of $2.30$ GHz.
Each processor has $18$ hardware cores and $2$ hyper-threads per core. The system has total shared memory of $128$ GB and is running Ubuntu $20.04.2$ LTS ($64$ bit) and Linux Kernel version $5.4.0$-$71$.

We used LLVM OpenMP runtime version $5.0$ for \tool{TSan-LLVM}, \tool{LLOV}, \tool{OMPRacer}, \tool{Archer}, and \tool{SWORD}. 
For \tool{Helgrind}, \tool{Valgrind DRD}, and \tool{ROMP}, we used gcc version $9.3.0$ and GOMP in our experiments.
clang $9.0.1$ was required for \tool{OMPRacer} and clang $12.0.1$ was used for the rest.

\textbf{Performance Metrics:}
\EM{True Positive} (TP): The outcome of a tool on a kernel with a known data race is considered as TP if the tool reports a race at least once on that kernel in all its runs.
If the tool fails to report a race even once on such a kernel then it is considered as \EM{False Negative} (FN).

\noindent \EM{True Negative} (TN): The outcome of a tool on a kernel with no data race is considered as TN if across all its runs on the kernel the tool reports no races. If the tool erroneously reports a race then it is considered a \EM{False Positive} (FP).

In addition, precision, recall, accuracy, F1 score, and diagnostics odds ratio are defined as follows:

\noindent \EM{Precision} = TP / (TP + FP)

\noindent \EM{Recall} = TP / (TP + FN)

\noindent \EM{Accuracy} = (TP + TN) / (TP + FP + TN + FN)

\noindent \EM{F1 Score} = (2 * Precision * Recall) / (Precision + Recall)

\noindent \EM{Diagnostics Odds Ratio (DOR)} = $\frac{LR+}{LR-}$
where, \\
\begin{minipage}{0.5\textwidth}
Positive Likelihood Ratio $(LR+) = \frac{TPR}{FPR}$ ,\\
True Positive Rate $(TPR) = \frac{TP}{TP\ +\ FN}$ ,\\
False Positive Rate $(FPR) = \frac{FP}{FP\ +\ TN}$ ,
\end{minipage}
\begin{minipage}{0.5\textwidth}
Negative Likelihood Ratio $(LR-) = \frac{FNR}{TNR}$ ,\\
False Negative Rate $(FNR) = \frac{FN}{FN\ +\ TP}$ , and \\
True Negative Rare $(TNR) = \frac{TN}{TN\ +\ FP}$.
\end{minipage}

\subsection{Performance Evaluation}
For performance evaluation, we have used DataRaceBench v$1.3.2$ consisting of kernels for which the absence or presence of a data race is known.
We compare our work against the state-of-the-art race detection tools \tool{TSan-LLVM}, \tool{Archer}, \tool{Valgrind DRD}, \tool{Helgrind}, \tool{ROMP}, \tool{SWORD}, and \tool{OMPRacer}.
The version numbers and flags used in our experiments are listed in~\tableref{tools}.

\begin{table*}[t]
\centering
\caption{Race detection tools with the version numbers and flags used for comparison}
\label{tbl:tools}
\resizebox{0.6\textwidth}{!}{%
\begin{tabular}{ lcl }
\toprule
\textbf{Tools} &
\textbf{Version / Commit} & \textbf{Flags} \\
\midrule
\tool{Helgrind}~\cite{valgrind2007helgrind} & 3.15.0 & \EM{-{}-tool=helgrind} \\
\tool{Valgrind DRD}~\cite{valgrind2007drd} & 3.15.0 & \EM{-{}-tool=drd -{}-check-stack-var=yes} \\
\tool{TSan-LLVM}~\cite{Serebryany:rv:2011} & 12.0.1 & \EM{ignore\_noninstrumented\_modules=1} \\
\tool{Archer}~\cite{Atzeni:ipdps:2016} & 5b37681 & \EM{ignore\_noninstrumented\_modules=1} \\
\tool{SWORD}~\cite{Atzeni:ipdps:2018} & 7a08f3c & \EM{-{}-analysis-tool=sword-race-analysis} \\
\tool{ROMP}~\cite{Gu:sc:2018} & b3e248e & \\
\tool{OMPRacer}~\cite{Swain:sc:2020} & 0.1.1 & \EM{-{}-no-filter -{}-silent -{}-nolimit} \\
\llovpolly~\cite{Bora:taco:2020} &  & \EM{openmp-verify-use-llvm-da=false} \\
\llovmhponly & & \EM{openmp-verify-use-polly-da=false} \\
\llovmhp &  & \EM{-Xclang -disable-O0-optnone} \\
\bottomrule
\end{tabular}
}
\end{table*}

\noindent\textbf{DataRaceBench v$1.3.2$}~\cite{Verma:Correctness:2020} is a seeded OpenMP benchmark designed for data race detection, and is being commonly used in the literature.

DataRaceBench is a collection of OpenMP programs making use of a variety of OpenMP features and pragmas. The programs exhibit common data race conditions due to various reasons, such as, missing data sharing clauses, input dependent data races, improper synchronizations, and data races that are dependent on number of runtime threads.
The benchmark (v$1.3.2$) consists of $172$ kernels out of which $89$ kernels do not have any data races and the remaining $83$ kernels have known data races. Most of the kernels with races have only one data race each. A few kernels, however, have more than one data race, either in the form of shared induction variables or scalar accesses to shared memory inside a loop.

We consider an outcome of a tool to be TP if it can detect at least one race in the kernels with more than one data races.

For the dynamic tools, our experiments use two parameters similar to Liao et al.~\cite{Liao:sc:2017} and Bora et al.~\cite{Bora:taco:2020}: (i) the number of OpenMP threads,  and (ii) the input size for variable length arrays.  
The number of threads that we considered for the experiments are $\{3,36,45,72,90,180,256\}$. 
For the 16 variable length kernels, we considered 6 different array sizes as follows: $\{32,64,128,256,512,1024\}$.
With each particular set of parameters, we ran each of the $172$ kernels $5$ times.
Both the number of threads and array sizes have been used in prior studies~\cite{Liao:sc:2017, Bora:taco:2020}, and we have used the same setup for uniformity.
The 16 kernels with variable length arrays were run $3360$ ( $=16$ kernels $\times$ $7$ thread sizes $\times$ $6$ array sizes $\times$ $5$ runs) times in total.
The remaining $156$ kernels were run $5460$ ($156$ kernels $\times$ $7$ thread sizes $\times$ $5$ runs ) times in total.
For the static tools, we run each kernel $5$ times.
For every compilation and run, we have used a timeout of $300$ seconds for compilation as well as execution separately.

\begin{table*}[t]
\caption{Maximum number of Races reported by different tools in DataRaceBench v1.3.2}
\centering
\small
\resizebox{.9\textwidth}{!}{
\begin{tabular}{ |l|cc|cc|c|c|c|c|c|c| }
 \hline
 \cline{1-11}
 \multirow{2}{*}{Tools} & \multicolumn{2}{c|}{Race: Yes} & \multicolumn{2}{c|}{Race: No} &
 Coverage & \multicolumn{5}{c|}{Performance Metrics} \\ 
 \cline{2-5}\cline{7-11} & TP & FN & TN & FP & (172) & Precision & Recall & Accuracy & F1 Score & DOR \\
 \hline
 \tool{Helgrind} & 67 & 4 & 7 & 68 & 146 & 0.50 & 0.94 & 0.51 & 0.65 & 1.72 \\
 \hline
 \tool{Valgrind DRD} & 67 & 4 & 33 & 42 & 146 & 0.61 & 0.94 & 0.68 & 0.74 & 13.16 \\
 \hline
 \tool{TSan-LLVM} & \textbf{74} & 6 & 48 & 38 & 166 & 0.66 & 0.93 & 0.73 & 0.77 & 15.58 \\
 \hline
 \tool{Archer} & 69 & 12 & \textbf{78} & 7 & 166 & 0.91 & 0.85 & 0.89 & 0.88 & 64.07 \\
 \hline
 \tool{SWORD} & 51 & 19 & 45 & 6 & 121 & 0.89 & 0.73 & 0.79 & 0.80 & 20.13 \\
 \hline
 \tool{ROMP} & 63 & 17 & 74 & 11 & 165 & 0.85 & 0.79 & 0.83 & 0.82 & 24.93 \\
 \hline
 \tool{OMPRacer} & 66 & 15 & 69 & 12 & 162 & 0.85 & 0.81 & 0.83 & 0.83 & 25.30 \\
 \hline
 \tool{OMPRacer}(notasks) & 65 & 6 & 49 & 11 & 131 & 0.86 & 0.92 & 0.87 & 0.88 & 48.26 \\
 \hline 
 \hline
 \llovpolly & 46 & 7 & 24 & 3 & 80 & \textbf{0.94} & 0.87 & 0.88 & 0.90 & 52.57 \\
 \hline
 \llovmhponly & 70 & 1 & 45 & 24 & 140 & 0.74 & \textbf{0.99} & 0.82 & 0.85 & 131.25 \\
 \hline
 \llovmhp & 70 & 1 & 60 & 8 & 139 & 0.90 & \textbf{0.99} & \textbf{0.94} & \textbf{0.94} & \textbf{525.00} \\
 \hline
\end{tabular}
}
\label{tbl:drbv13combined}
\end{table*}
%=============================

In~\tableref{drbv13combined}, column ``TP" represents the number of True Positive results reported by a tool.
Similarly, columns ``FN", ``TN", and ``FP" represents False Negative, True Negative, and False Positive outcomes respectively. 
Coverage out of $172$ kernels is shown in column ``Coverage".
The performance metrics for each tool is also shown in a similar fashion.
The best numbers are highlighted in bold.

\tool{OMPRacer} does not support tasking constructs, and there is no option to know at runtime when a pragma is not supported.
Hence, we have also added the numbers for \tool{OMPRacer} (notasks) by excluding the kernels that have tasking constructs.
On the other hand, \llovmhp explicitly flags when a pragma is not supported. Also, kernels with unhandled pragmas are not taken into account towards its coverage.

It can be seen from~\tableref{drbv13combined} that \llovmhp performs the best in terms of Recall, Accuracy, F1 Score, and DOR.
\llovpolly has the best Precision, followed by \tool{Archer} and \llovmhp closely in third position. \\

\subsection{Scalability Analysis}

\EM{ECP Proxy applications: } 
The Exascale Computing Project (ECP) is a push towards developing an exascale ecosystem for scientific computing in the United States.
As part of the ECP project, multiple mini proxy applications were developed to design and test new programming models, technologies, and architectures for supercomputing.
We have used the proxy apps AMG, miniAMR, miniQMC, miniVite, SW4lite, and XSBench from the ECP Proxy Apps Suite Release $4.0$ which have implementations of parallel algorithms using OpenMP.\\
Other significant proxy applications includes CoMD, LULESH, miniBUDE, miniFE, and RSBench.
Proxy applications and their sizes in lines of code are listed in~\tableref{misc-bench-loc}.

\begin{figure}[t]
\begin{tikzpicture}
\centering
\begin{axis}[
  scale=.85,
  xlabel=Number of Benchmarks,
  ylabel=Time in Seconds,
  xticklabel=\pgfmathprintnumber{\tick},
  xticklabel style={
    rotate=90,
    font=\scriptsize,
    },
  every node near coord/.append style={
    anchor=east,
    rotate=90
  },
  xtick=data,
  scaled x ticks = true,
  legend pos=outer north east,
  legend plot pos=left,
  legend cell align=left, 
  legend columns = 2,
  legend style={ 
    nodes={scale=0.8,},
    column sep=1ex,
    at={(0.3,0.95)},
	anchor=north,
    font=\scriptsize,
  },
  title=Cactus plot of different race detection tools,
]
\addplot+[
  color=red,
  fill=none,
  mark=triangle,
 ]
table [
  y = Time,
  x = NumBenchmarks
  ] {./Data/Misc_Archer_T_cactus.txt};
\addlegendentry{\tool{Archer}}

\addplot+[
  color=green,
  fill=none,
  mark=halfsquare left*,
 ]
table [
  y = Time,
  x = NumBenchmarks
  ] {./Data/Misc_Helgrind_T_cactus.txt};
\addlegendentry{\tool{Helgrind}}

\addplot+[
  color=orange,
  fill=none,
  mark=triangle*,
]
table [
  y = Time,
  x = NumBenchmarks
  ] {./Data/Misc_TSan_T_cactus.txt};
\addlegendentry{\tool{TSan}}

\addplot+[
  color=gray,
  fill=none,
  mark=halfsquare right*,
 ]
table [
  y = Time,
  x = NumBenchmarks
  ] {./Data/Misc_DRD_T_cactus.txt};
\addlegendentry{\tool{DRD}}

\addplot+[
  color=cyan,
  fill=none,
  mark=halfdiamond*,
 ]
table [
  y = Time,
  x = NumBenchmarks
  ] {./Data/Misc_ROMP_T_cactus.txt};
\addlegendentry{\tool{ROMP}}

\addplot+[
  color=black,
  fill=none,
  mark=otimes,
 ]
table [
  y = Time,
  x = NumBenchmarks
  ] {./Data/Misc_OMPRacer_T_cactus.txt};
\addlegendentry{\tool{OMPRacer}}

\addplot+[
  color=teal,
  fill=none,
  mark=star,
 ]
table [
  y = Time,
  x = NumBenchmarks
  ] {./Data/Misc_LLOV_T_cactus.txt};
\addlegendentry{\tool{\toolname}}
\end{axis}
\end{tikzpicture}
\caption{Cactus plot of different race detection tools on miscellaneous benchmarks listed in~\tableref{misc-bench-loc}. X axis represents number of benchmarks analyzed, Y axis represents cumulative time.}
\label{fig:miscBench_time_cactus}
\end{figure}
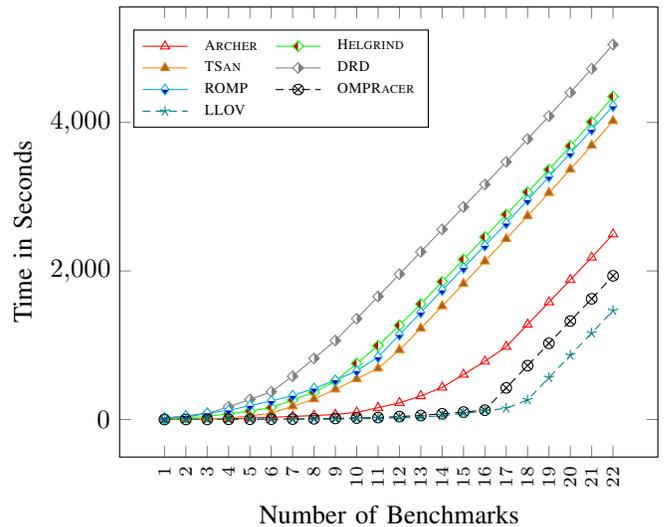

\noindent\EM{Large Applications: }
We evaluated \llovmhp on multiple large OpenMP applications with thousands of lines of code such as COVID-SIM, Sundials solver, Rodinia, and Parallel Research Kernels as listed in~\tableref{misc-bench-loc}.

Each of the tools reported combined thousands of data races on the $22$ benchmarks.
Since these applications have total lines of code exceeding a million, and they are not annotated with known data races, we have not yet validated the True Positive cases.
Rather, we analyze the tools based on the time taken to analyze them and their memory requirements.
Each of the $22$ benchmarks is executed $3$ times for each of the race detection tools using $32$ OpenMP threads and default program parameters specified in the benchmarks.
Timeout of $300$ seconds was used for compilation and execution separately.
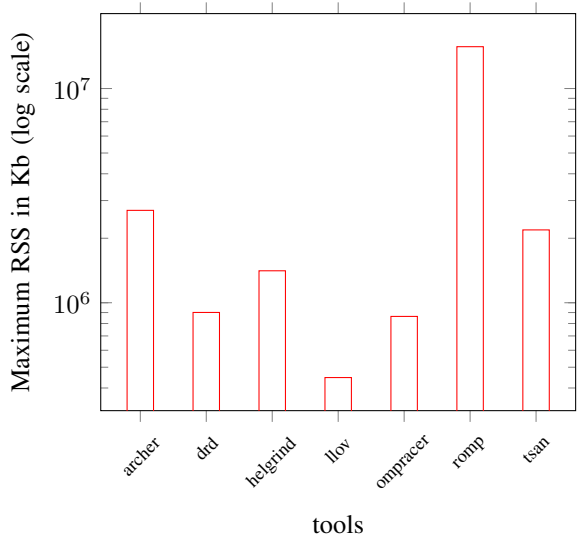
\begin{figure}[t]
\begin{tikzpicture}
\centering
\begin{axis}[
  scale=.75,
  ybar,
  bar width=10pt,
  xlabel=tools,
  ylabel=Maximum RSS in Kb (log scale),
  ymode=log,
  log basis y={10},
  xticklabel style={
    rotate=45,
    font=\scriptsize,
    },
  every node near coord/.append style={
    anchor=east,
    rotate=45
  },
  xtick=data,
  xticklabels from table={./Data/Misc_averave_memory.txt}{tools},
  scaled x ticks = true,
  legend pos=outer north east,
  legend plot pos=left,
  legend cell align=left, 
  legend columns = 2,
  legend style={ 
    nodes={scale=0.8,},
    column sep=1ex,
    at={(0.7,0.25)},
	anchor=north,
    font=\scriptsize,
  },
  title=Average maximum RSS of the tools across all the benchmarks,
]
\addplot+[
  color=red,
  fill=none,
 ]
table [
  %x = tools,
  x expr=\coordindex,
  y = memory
  ] {./Data/Misc_averave_memory.txt};

\end{axis}
\end{tikzpicture}
\caption{Average maximum resident set size of the tools across all the $22$ benchmarks listed in~\tableref{misc-bench-loc}}
\label{fig:miscBench_mem_avg}
\end{figure}

~\figref{miscBench_time_cactus} shows a plot of time taken by various tools for the $22$ benchmarks listed in~\tableref{misc-bench-loc}.
Horizontal axis represents the number of benchmarks processed by a tool and vertical axis represents the cumulative time taken (compiletime and runtime for dynamic tools and only compiletime for static tools) to process all the benchmarks.
A tool is penalized by the timeout value of $300$ seconds if it fails to analyze a benchmark due to compilation error or runtime failure.
The  side-effect of the timeout/penalty is visible in the plot as one can see a linear increase in time after a certain point for each tool.

Note that 
\llovmhp scaled exceedingly well with the real world applications and could analyze them, on average, in less than $60$ seconds including penalties.
It can be seen that \llovmhp could process maximum number of benchmarks in the least amount of time, closely followed by \tool{OMPRacer}.

~\figref{miscBench_mem_avg} shows the arithmetic mean of the maximum resident set size (RSS) depicting the memory consumption for each of the tools.
The memory requirement for \llovmhp is significantly low compared to the other tools.

\begin{table}[t]
\caption{Number of Lines of Code in $22$ Miscellaneous Benchmarks}
\centering
\resizebox{\columnwidth}{!}{
\begin{tabular}{ |l|c|l|c| }
 \hline
\textbf{Benchmark} & \textbf{Lines of Code} & \textbf{Benchmark} & \textbf{Lines of Code} \\
\hline
AMG & 64.9k & QuickSilver & 10k \\
CoMD & 6.1k & RSBench & 6k \\
COVID-SIM & 19k & Sundials & 186.9k \\
LCALS & 6.6k & sw4lite & 54.2k \\
LULESH & 5.5k & TriangleCounting & .8k \\
miniAMR & 17.7k & XSBench & 6.7k \\
miniBUDE & 9.5k & Rodinia & 186.4k \\
miniFE & 297.8k & ParallelResearchKernels & 106.2k \\
miniQMC & 31.9k & OpenMP-Microbench & 1k \\
miniVite & 2.4k & openmp-tutorial & 10.4k \\
NBody & 0.4k & NAS-Parallel & 15.9k \\
\hline
\end{tabular}
}
\label{tbl:misc-bench-loc}
\end{table}
%=============================

\section{Summary and Future work}\label{Se:futurework}

We propose a May-Happen-in-Parallel (MHP) based analysis for data race checking of parallel programs written in OpenMP. Our static analysis based data race checker is based on Phase Interval Analysis (PIA).

Through our experiments it is established that while using MHP analysis increases the coverage, it is the combination of MHP as well as the exact dependence analysis provided by Polly which provides the 
best results in terms of accuracy. 
We have compared \llovmhp  with several other state-of-the-art tools, and the results show that this combination provides for much superior  overall accuracy.

We also proved the improved scalability of our tool on code bases with a combined lines of code exceeding a million. \llovmhp is not only significantly faster on big benchmarks but also has a very small memory footprint as compared to other race-checkers.

In future, we want to improve the coverage of the pragmas by adding support for tasking constructs.
We would like to explore using PIA in compiler optimization for OpenMP programs.
\section*{Acknowledgment}

This research is funded by the Department of Electronics \& Information Technology and the Ministry of Communications \& Information Technology, Government of India. This work is partially supported by a Visvesvaraya PhD Scheme under the MEITY, GoI (PhD-MLA/04(02)/2015-16), an NSM research grant (MeitY/R\&D/HPC/2(1)/2014), a Visvesvaraya Young Faculty Research Fellowship from MeitY, and a faculty research grant from AMD.
\bibliographystyle{IEEEtran}
\bibliography{Common/references.bib,Common/MHP.bib}

\begin{thebibliography}{10}
\providecommand{\url}[1]{#1}
\csname url@samestyle\endcsname
\providecommand{\newblock}{\relax}
\providecommand{\bibinfo}[2]{#2}
\providecommand{\BIBentrySTDinterwordspacing}{\spaceskip=0pt\relax}
\providecommand{\BIBentryALTinterwordstretchfactor}{4}
\providecommand{\BIBentryALTinterwordspacing}{\spaceskip=\fontdimen2\font plus
\BIBentryALTinterwordstretchfactor\fontdimen3\font minus
  \fontdimen4\font\relax}
\providecommand{\BIBforeignlanguage}[2]{{%
\expandafter\ifx\csname l@#1\endcsname\relax
\typeout{** WARNING: IEEEtran.bst: No hyphenation pattern has been}%
\typeout{** loaded for the language `#1'. Using the pattern for}%
\typeout{** the default language instead.}%
\else
\language=\csname l@#1\endcsname
\fi
#2}}
\providecommand{\BIBdecl}{\relax}
\BIBdecl

\bibitem{top500url}
TOP500.Org, ``{Top 500 Supercomputer Sites},''
  \url{https://top500.org/lists/top500/2021/06/}, 2021, [Online; accessed
  28-June-2021].

\bibitem{valgrind2007helgrind}
Valgrind-project, ``{Helgrind: a thread error detector},''
  \url{http://valgrind.org/docs/manual/hg-manual.html}, 2007, [Online; accessed
  08-May-2019].

\bibitem{valgrind2007drd}
------, ``{DRD: a thread error detector},''
  \url{http://valgrind.org/docs/manual/drd-manual.html}, 2007, [Online;
  accessed 08-May-2019].

\bibitem{Serebryany:wbia:2009}
\BIBentryALTinterwordspacing
K.~Serebryany and T.~Iskhodzhanov, ``Threadsanitizer: Data race detection in
  practice,'' in \emph{Proceedings of the Workshop on Binary Instrumentation
  and Applications}, ser. WBIA '09.\hskip 1em plus 0.5em minus 0.4em\relax New
  York, NY, USA: ACM, 2009, pp. 62--71. [Online]. Available:
  \url{http://doi.acm.org/10.1145/1791194.1791203}
\BIBentrySTDinterwordspacing

\bibitem{Atzeni:ipdps:2016}
\BIBentryALTinterwordspacing
S.~Atzeni, G.~Gopalakrishnan, Z.~Rakamari\'{c}, D.~H. Ahn, I.~Laguna,
  M.~Schulz, G.~L. Lee, J.~Protze, and M.~S. M{\"{u}}ller, ``{ARCHER:}
  effectively spotting data races in large openmp applications,'' in \emph{2016
  {IEEE} International Parallel and Distributed Processing Symposium, {IPDPS}
  2016, Chicago, IL, USA, May 23-27, 2016}.\hskip 1em plus 0.5em minus
  0.4em\relax {IEEE} Computer Society, 2016, pp. 53--62. [Online]. Available:
  \url{https://doi.org/10.1109/IPDPS.2016.68}
\BIBentrySTDinterwordspacing

\bibitem{Bora:taco:2020}
\BIBentryALTinterwordspacing
U.~Bora, S.~Das, P.~Kukreja, S.~Joshi, R.~Upadrasta, and S.~Rajopadhye,
  ``{LLOV: A Fast Static Data-Race Checker for OpenMP Programs},'' \emph{ACM
  Trans. Archit. Code Optim.}, vol.~17, no.~4, Dec. 2020. [Online]. Available:
  \url{https://doi.org/10.1145/3418597}
\BIBentrySTDinterwordspacing

\bibitem{Basupalli:iwomp:2011}
\BIBentryALTinterwordspacing
V.~Basupalli, T.~Yuki, S.~Rajopadhye, A.~Morvan, S.~Derrien, P.~Quinton, and
  D.~Wonnacott, ``ompverify: Polyhedral analysis for the openmp programmer,''
  in \emph{OpenMP in the Petascale Era - 7th International Workshop on OpenMP,
  {IWOMP} 2011, Chicago, IL, USA, June 13-15, 2011. Proceedings}, ser. Lecture
  Notes in Computer Science, B.~M. Chapman, W.~D. Gropp, K.~Kumaran, and M.~S.
  M{\"{u}}ller, Eds., vol. 6665.\hskip 1em plus 0.5em minus 0.4em\relax Berlin,
  Heidelberg: Springer Berlin Heidelberg, 2011, pp. 37--53. [Online].
  Available: \url{https://doi.org/10.1007/978-3-642-21487-5\_4}
\BIBentrySTDinterwordspacing

\bibitem{Chatarasi:lcpc:2016}
\BIBentryALTinterwordspacing
P.~Chatarasi, J.~Shirako, M.~Kong, and V.~Sarkar, ``An extended polyhedral
  model for {SPMD} programs and its use in static data race detection,'' in
  \emph{Languages and Compilers for Parallel Computing - 29th International
  Workshop, {LCPC} 2016, Rochester, NY, USA, September 28-30, 2016, Revised
  Papers}, ser. Lecture Notes in Computer Science, C.~Ding, J.~Criswell, and
  P.~Wu, Eds., vol. 10136.\hskip 1em plus 0.5em minus 0.4em\relax Springer,
  2016, pp. 106--120. [Online]. Available:
  \url{https://doi.org/10.1007/978-3-319-52709-3\_10}
\BIBentrySTDinterwordspacing

\bibitem{Ye:correctness:2018}
\BIBentryALTinterwordspacing
F.~Ye, M.~Schordan, C.~Liao, P.~Lin, I.~Karlin, and V.~Sarkar, ``Using
  polyhedral analysis to verify openmp applications are data race free,'' in
  \emph{2nd {IEEE/ACM} International Workshop on Software Correctness for {HPC}
  Applications, CORRECTNESS@SC 2018, Dallas, TX, USA, November 12, 2018},
  I.~Laguna and C.~Rubio{-}Gonz{\'{a}}lez, Eds.\hskip 1em plus 0.5em minus
  0.4em\relax {IEEE}, 2018, pp. 42--50. [Online]. Available:
  \url{https://doi.org/10.1109/Correctness.2018.00010}
\BIBentrySTDinterwordspacing

\bibitem{Lattner:lcpc:2004}
C.~Lattner and V.~Adve, ``{The LLVM Compiler Framework and Infrastructure
  Tutorial},'' in \emph{{LCPC'04 Mini Workshop on Compiler Research
  Infrastructures}}, West Lafayette, Indiana, Sep 2004.

\bibitem{Swain:sc:2020}
B.~Swain, Y.~Li, P.~Liu, I.~Laguna, G.~Georgakoudis, and J.~Huang, ``{OMPRacer:
  A Scalable and Precise Static Race Detector for OpenMP Programs},'' in
  \emph{Proceedings of the International Conference for High Performance
  Computing, Networking, Storage and Analysis}, ser. SC '20.\hskip 1em plus
  0.5em minus 0.4em\relax IEEE Press, 2020.

\bibitem{Bernstein:TEC:1966}
A.~J. Bernstein, ``Analysis of programs for parallel processing,'' \emph{IEEE
  Transactions on Electronic Computers}, vol. EC-15, no.~5, pp. 757--763, 1966.

\bibitem{Feautrier:Encyclopedia:2011}
\BIBentryALTinterwordspacing
P.~Feautrier and C.~Lengauer, \emph{Polyhedron Model}.\hskip 1em plus 0.5em
  minus 0.4em\relax Boston, MA: Springer US, 2011, pp. 1581--1592. [Online].
  Available: \url{https://doi.org/10.1007/978-0-387-09766-4\_502}
\BIBentrySTDinterwordspacing

\bibitem{Grosser:ppl:2012}
\BIBentryALTinterwordspacing
T.~Grosser, A.~Gr{\"o}{\ss}linger, and C.~Lengauer, ``Polly -- performing
  polyhedral optimizations on a low-level intermediate representation,''
  \emph{Parallel Processing Letters}, vol.~22, no.~04, 2012. [Online].
  Available:
  \url{http://www.worldscientific.com/doi/abs/10.1142/S0129626412500107}
\BIBentrySTDinterwordspacing

\bibitem{Lattner:cgo:2004}
\BIBentryALTinterwordspacing
C.~Lattner and V.~Adve, ``{LLVM: A Compilation Framework for Lifelong Program
  Analysis \& Transformation},'' in \emph{Proceedings of the International
  Symposium on Code Generation and Optimization: Feedback-directed and Runtime
  Optimization}, ser. CGO '04.\hskip 1em plus 0.5em minus 0.4em\relax
  Washington, DC, USA: IEEE Computer Society, 2004, pp. 75--. [Online].
  Available: \url{http://dl.acm.org/citation.cfm?id=977395.977673}
\BIBentrySTDinterwordspacing

\bibitem{Gu:sc:2018}
Y.~Gu and J.~Mellor-Crummey, ``Dynamic data race detection for openmp
  programs,'' in \emph{Proceedings of the International Conference for High
  Performance Computing, Networking, Storage, and Analysis}, ser. SC '18.\hskip
  1em plus 0.5em minus 0.4em\relax IEEE Press, 2018.

\bibitem{Barik:lcpc:2005}
\BIBentryALTinterwordspacing
R.~Barik, ``Efficient computation of may-happen-in-parallel information for
  concurrent java programs,'' in \emph{Languages and Compilers for Parallel
  Computing, 18th International Workshop, {LCPC} 2005, Hawthorne, NY, USA,
  October 20-22, 2005, Revised Selected Papers}, ser. Lecture Notes in Computer
  Science, E.~Ayguad{\'{e}}, G.~Baumgartner, J.~Ramanujam, and P.~Sadayappan,
  Eds., vol. 4339.\hskip 1em plus 0.5em minus 0.4em\relax Springer, 2005, pp.
  152--169. [Online]. Available:
  \url{https://doi.org/10.1007/978-3-540-69330-7\_11}
\BIBentrySTDinterwordspacing

\bibitem{Joshi:ipdpsw:2012}
S.~Joshi, R.~K. Shyamasundar, and S.~K. Aggarwal, ``{A New Method of MHP
  Analysis for Languages with Dynamic Barriers},'' in \emph{2012 IEEE 26th
  International Parallel and Distributed Processing Symposium Workshops \& PhD
  Forum}, 2012, pp. 519--528.

\bibitem{Cousot:ISP:1976}
P.~Cousot and R.~Cousot, ``Static determination of dynamic properties of
  programs,'' in \emph{Proceedings of the Second International Symposium on
  Programming}.\hskip 1em plus 0.5em minus 0.4em\relax Dunod, Paris, France,
  1976, pp. 106--130.

\bibitem{Savage:tocs:1997}
\BIBentryALTinterwordspacing
S.~Savage, M.~Burrows, G.~Nelson, P.~Sobalvarro, and T.~Anderson, ``Eraser: A
  dynamic data race detector for multithreaded programs,'' \emph{ACM Trans.
  Comput. Syst.}, vol.~15, no.~4, pp. 391--411, Nov. 1997. [Online]. Available:
  \url{https://doi.org/10.1145/265924.265927}
\BIBentrySTDinterwordspacing

\bibitem{Naumovich:fse:1998}
\BIBentryALTinterwordspacing
G.~Naumovich and G.~S. Avrunin, ``A conservative data flow algorithm for
  detecting all pairs of statements that may happen in parallel,'' in
  \emph{Proceedings of the 6th ACM SIGSOFT International Symposium on
  Foundations of Software Engineering}, ser. SIGSOFT '98/FSE-6.\hskip 1em plus
  0.5em minus 0.4em\relax New York, NY, USA: Association for Computing
  Machinery, 1998, p. 24–34. [Online]. Available:
  \url{https://doi.org/10.1145/288195.288213}
\BIBentrySTDinterwordspacing

\bibitem{Naumovich:fse:1999}
G.~Naumovich, G.~S. Avrunin, and L.~A. Clarke, ``An efficient algorithm for
  computing <i>mhp</i> information for concurrent java programs,'' in
  \emph{Proceedings of the 7th European Software Engineering Conference Held
  Jointly with the 7th ACM SIGSOFT International Symposium on Foundations of
  Software Engineering}, ser. ESEC/FSE-7.\hskip 1em plus 0.5em minus
  0.4em\relax Berlin, Heidelberg: Springer-Verlag, 1999, p. 338–354.

\bibitem{Agarwal:ppopp:2007}
\BIBentryALTinterwordspacing
S.~Agarwal, R.~Barik, V.~Sarkar, and R.~K. Shyamasundar,
  ``May-happen-in-parallel analysis of x10 programs,'' in \emph{Proceedings of
  the 12th ACM SIGPLAN Symposium on Principles and Practice of Parallel
  Programming}, ser. PPoPP '07.\hskip 1em plus 0.5em minus 0.4em\relax New
  York, NY, USA: Association for Computing Machinery, 2007, p. 183–193.
  [Online]. Available: \url{https://doi.org/10.1145/1229428.1229471}
\BIBentrySTDinterwordspacing

\bibitem{Saha:ppopp:2020}
\BIBentryALTinterwordspacing
S.~Saha and V.~K. Nandivada, ``On the fly mhp analysis,'' in \emph{Proceedings
  of the 25th ACM SIGPLAN Symposium on Principles and Practice of Parallel
  Programming}, ser. PPoPP '20.\hskip 1em plus 0.5em minus 0.4em\relax New
  York, NY, USA: Association for Computing Machinery, 2020, p. 173–186.
  [Online]. Available: \url{https://doi.org/10.1145/3332466.3374541}
\BIBentrySTDinterwordspacing

\bibitem{Sankar:cc:2016}
\BIBentryALTinterwordspacing
A.~Sankar, S.~Chakraborty, and V.~K. Nandivada, ``Improved mhp analysis,'' in
  \emph{Proceedings of the 25th International Conference on Compiler
  Construction}, ser. CC 2016.\hskip 1em plus 0.5em minus 0.4em\relax New York,
  NY, USA: Association for Computing Machinery, 2016, p. 207–217. [Online].
  Available: \url{https://doi.org/10.1145/2892208.2897144}
\BIBentrySTDinterwordspacing

\bibitem{Engler:sosp:2003}
\BIBentryALTinterwordspacing
D.~Engler and K.~Ashcraft, ``Racerx: Effective, static detection of race
  conditions and deadlocks,'' in \emph{Proceedings of the Nineteenth ACM
  Symposium on Operating Systems Principles}, ser. SOSP '03.\hskip 1em plus
  0.5em minus 0.4em\relax New York, NY, USA: Association for Computing
  Machinery, 2003, pp. 237--252. [Online]. Available:
  \url{https://doi.org/10.1145/945445.945468}
\BIBentrySTDinterwordspacing

\bibitem{Yu:sosp:2005}
\BIBentryALTinterwordspacing
Y.~Yu, T.~Rodeheffer, and W.~Chen, ``Racetrack: Efficient detection of data
  race conditions via adaptive tracking,'' in \emph{Proceedings of the
  Twentieth ACM Symposium on Operating Systems Principles}, ser. SOSP
  '05.\hskip 1em plus 0.5em minus 0.4em\relax New York, NY, USA: Association
  for Computing Machinery, 2005, pp. 221--234. [Online]. Available:
  \url{https://doi.org/10.1145/1095810.1095832}
\BIBentrySTDinterwordspacing

\bibitem{Kasikci:hotdep:2012}
B.~Kasikci, C.~Zamfir, and G.~Candea, ``{CoRD: A Collaborative Framework for
  Distributed Data Race Detection},'' in \emph{Proceedings of the Eighth USENIX
  Conference on Hot Topics in System Dependability}, ser. HotDep'12.\hskip 1em
  plus 0.5em minus 0.4em\relax USA: USENIX Association, 2012, p.~4.

\bibitem{Voung:fse:2007}
\BIBentryALTinterwordspacing
J.~W. Voung, R.~Jhala, and S.~Lerner, ``Relay: Static race detection on
  millions of lines of code,'' in \emph{Proceedings of the the 6th Joint
  Meeting of the European Software Engineering Conference and the ACM SIGSOFT
  Symposium on The Foundations of Software Engineering}, ser. ESEC-FSE
  '07.\hskip 1em plus 0.5em minus 0.4em\relax New York, NY, USA: Association
  for Computing Machinery, 2007, pp. 205--214. [Online]. Available:
  \url{https://doi.org/10.1145/1287624.1287654}
\BIBentrySTDinterwordspacing

\bibitem{Pratikakis:pldi:2006}
\BIBentryALTinterwordspacing
P.~Pratikakis, J.~S. Foster, and M.~Hicks, ``Locksmith: Context-sensitive
  correlation analysis for race detection,'' in \emph{Proceedings of the 27th
  ACM SIGPLAN Conference on Programming Language Design and Implementation},
  ser. PLDI '06.\hskip 1em plus 0.5em minus 0.4em\relax New York, NY, USA:
  Association for Computing Machinery, 2006, p. 320–331. [Online]. Available:
  \url{https://doi.org/10.1145/1133981.1134019}
\BIBentrySTDinterwordspacing

\bibitem{Lamport:cacm:1978}
\BIBentryALTinterwordspacing
L.~Lamport, ``Time, clocks, and the ordering of events in a distributed
  system,'' \emph{Commun. ACM}, vol.~21, no.~7, pp. 558--565, Jul. 1978.
  [Online]. Available: \url{http://doi.acm.org/10.1145/359545.359563}
\BIBentrySTDinterwordspacing

\bibitem{Flanagan:pldi:2009}
\BIBentryALTinterwordspacing
C.~Flanagan and S.~N. Freund, ``Fasttrack: Efficient and precise dynamic race
  detection,'' in \emph{Proceedings of the 30th ACM SIGPLAN Conference on
  Programming Language Design and Implementation}, ser. PLDI '09.\hskip 1em
  plus 0.5em minus 0.4em\relax New York, NY, USA: Association for Computing
  Machinery, 2009, p. 121–133. [Online]. Available:
  \url{https://doi.org/10.1145/1542476.1542490}
\BIBentrySTDinterwordspacing

\bibitem{Chatarasi:impact:2016}
P.~Chatarasi, J.~Shirako, and V.~Sarkar, ``Static data race detection for spmd
  programs via an extended polyhedral representation,'' in \emph{Proceedings of
  the 6th International Workshop on Polyhedral Compilation Techniques
  (IMPACT)}, 2016.

\bibitem{Mellor-Crummey:sc:1991}
\BIBentryALTinterwordspacing
J.~Mellor-Crummey, ``On-the-fly detection of data races for programs with
  nested fork-join parallelism,'' in \emph{Proceedings of the 1991 ACM/IEEE
  Conference on Supercomputing}, ser. Supercomputing '91.\hskip 1em plus 0.5em
  minus 0.4em\relax New York, NY, USA: Association for Computing Machinery,
  1991, pp. 24--33. [Online]. Available:
  \url{https://doi.org/10.1145/125826.125861}
\BIBentrySTDinterwordspacing

\bibitem{posix2017}
P.~A. J.~W. Group, ``Ieee standard for information technology--portable
  operating system interface (posix(r)) base specifications, issue 7,''
  \emph{IEEE Std 1003.1-2017 (Revision of IEEE Std 1003.1-2008)}, pp. 1--3951,
  Jan 2018.

\bibitem{Atzeni:ipdps:2018}
\BIBentryALTinterwordspacing
S.~Atzeni, G.~Gopalakrishnan, Z.~Rakamari\'{c}, I.~Laguna, G.~L. Lee, and D.~H.
  Ahn, ``{SWORD:} {A} bounded memory-overhead detector of openmp data races in
  production runs,'' in \emph{2018 {IEEE} International Parallel and
  Distributed Processing Symposium, {IPDPS} 2018, Vancouver, BC, Canada, May
  21-25, 2018}.\hskip 1em plus 0.5em minus 0.4em\relax {IEEE} Computer Society,
  2018, pp. 845--854. [Online]. Available:
  \url{https://doi.org/10.1109/IPDPS.2018.00094}
\BIBentrySTDinterwordspacing

\bibitem{Serebryany:rv:2011}
\BIBentryALTinterwordspacing
K.~Serebryany, A.~Potapenko, T.~Iskhodzhanov, and D.~Vyukov, ``Dynamic race
  detection with llvm compiler,'' in \emph{Proceedings of the Second
  International Conference on Runtime Verification}, ser. RV'11.\hskip 1em plus
  0.5em minus 0.4em\relax Berlin, Heidelberg: Springer-Verlag, 2011, pp.
  110--114. [Online]. Available:
  \url{https://doi.org/10.1007/978-3-642-29860-8\_9}
\BIBentrySTDinterwordspacing

\bibitem{Eichenberger:iwomp:2013}
A.~E. Eichenberger, J.~Mellor-Crummey, M.~Schulz, M.~Wong, N.~Copty,
  R.~Dietrich, X.~Liu, E.~Loh, and D.~Lorenz, ``Ompt: An openmp tools
  application programming interface for performance analysis,'' in \emph{OpenMP
  in the Era of Low Power Devices and Accelerators}, A.~P. Rendell, B.~M.
  Chapman, and M.~S. M{\"u}ller, Eds.\hskip 1em plus 0.5em minus 0.4em\relax
  Berlin, Heidelberg: Springer Berlin Heidelberg, 2013, pp. 171--185.

\bibitem{valgrind2003url}
N.~Nethercote and J.~Seward, ``{Valgrind: A program supervision framework},''
  \url{http://valgrind.org/}, 2003, [Online; accessed 08-May-2019].

\bibitem{Pratikakis:toplas:2011}
\BIBentryALTinterwordspacing
P.~Pratikakis, J.~S. Foster, and M.~Hicks, ``Locksmith: Practical static race
  detection for c,'' \emph{ACM Trans. Program. Lang. Syst.}, vol.~33, no.~1,
  Jan. 2011. [Online]. Available: \url{https://doi.org/10.1145/1889997.1890000}
\BIBentrySTDinterwordspacing

\bibitem{Verma:Correctness:2020}
G.~Verma, Y.~Shi, C.~Liao, B.~Chapman, and Y.~Yan, ``Enhancing dataracebench
  for evaluating data race detection tools,'' in \emph{2020 IEEE/ACM 4th
  International Workshop on Software Correctness for HPC Applications
  (Correctness)}, 2020, pp. 20--30.

\bibitem{Liao:sc:2017}
\BIBentryALTinterwordspacing
C.~Liao, P.-H. Lin, J.~Asplund, M.~Schordan, and I.~Karlin, ``Dataracebench: A
  benchmark suite for systematic evaluation of data race detection tools,'' in
  \emph{Proceedings of the International Conference for High Performance
  Computing, Networking, Storage and Analysis}, ser. SC '17.\hskip 1em plus
  0.5em minus 0.4em\relax New York, NY, USA: ACM, 2017, pp. 11:1--11:14.
  [Online]. Available: \url{http://doi.acm.org/10.1145/3126908.3126958}
\BIBentrySTDinterwordspacing

\end{thebibliography}

\end{document}